\documentclass[reprint,aps]{revtex4-1}
\usepackage{graphicx,float,setspace,mathtools,multirow}
\pdfoutput=1
\DeclareMathOperator{\sech}{sech}

\begin{document}
\title{Half-vicinity model and a phase diagram \\
for quantum oscillations in confined and degenerate Fermi gases}
\author{Alhun Aydin}
\author{Altug Sisman}
 \email{Corresponding author, sismanal@itu.edu.tr}
\affiliation{Nano Energy Research Group, Energy Institute, Istanbul Technical University, 34469, Istanbul, Turkey}
\date{\today}
\begin{abstract}
We propose an analytical model for the accurate calculation of size and density dependent quantum oscillations in thermodynamic and transport properties of confined and degenerate non-interacting Fermi gases. We provide a universal, material independent, recipe that explicitly separates oscillatory quantum regime from stationary classical regime. Our model quite accurately estimates quantum oscillations depending on confinement and degeneracy. We construct a phase diagram representing stationary and oscillatory regimes on degeneracy-confinement space. Analytical expressions of phase transition interfaces are derived for different dimensions. The critical point on the phase diagram, which separates entirely stationary and entirely oscillatory regions, is determined and their aspect ratio dependencies are examined. Quantum oscillations as well as their periods are analytically expressed for one-dimensional case. Accuracy of our model is verified through quantum oscillations in electronic specific heat capacity. We also compare the predictions of our half-vicinity model, based on bounded sums, with those of infinite sums, for the oscillatory violation of entropy-heat capacity equivalence in degenerate limit to show the accuracy of our model. Furthermore, similarities between functional behaviors of total occupancy variance and conventional density of states functions at Fermi level are discussed.
\end{abstract}
\maketitle

%%%        %%%%%%%%%%%%%%%%%%%%%%%%%%%%%%%%%%%%%%%%%%%%%%%%%%%%%%%%%%%        %%%
%%%        %%%%%%%%%%%%%%%%%%%%%%%% SECTION 1 %%%%%%%%%%%%%%%%%%%%%%%%        %%%
%%%        %%%%%%%%%%%%%%%%%%%%%%%%%%%%%%%%%%%%%%%%%%%%%%%%%%%%%%%%%%%        %%%

\section{Introduction}
A great deal of attention has been given to the physics of low-dimensional nanostructures associated with the intensive developments of nanoscience and nanotechnology in recent years. When the mean de Broglie wavelength of particles inside a domain becomes comparable to the size of the domain, quantum size effects start to reveal themselves and their presence has to be taken into account in the calculation of the physical properties of such nanoscale systems \cite{mitchen,qbook,pathbook,broadbook}.

In Fermi gases, instead of fully occupied states, only partially occupied ones around Fermi level contribute to some thermodynamic quantities like entropy and specific heat and transport properties like electrical and thermal conductivities. Definitional expressions of such quantities contain the variance of distribution function instead of the distribution function itself. Hence, variance function plays a crucial role in the behavior of these quantities. In degenerate and strongly confined ideal Fermi gases, the quantities containing variance function exhibit oscillatory behaviors. The basic reason of this behavior is the oscillatory behavior of total occupancy variance (TOV) function due to changes in size and density.

It has been shown experimentally in literature that, size and density dependent oscillations occur in some certain thermodynamic and transport properties of charge carriers in semiconductors/metals or Fermi gases confined at nano domains in general. First experimental studies of size dependent oscillations due to quantum size effects has been reported in late $60$'s \cite{first1,first2,first3,first4}. Thermodynamic properties like entropy and specific heat capacity at constant volume, thermoelectric properties such as Seebeck coefficient and electronic transport properties such as charge carrier mobility, electrical and thermal conductivities are some examples of these oscillatory quantities \cite{broadbook,natosfer,eltrp,sbarma,heatco,PhysRevB.9.3347,anewt,qsedressel1,qsedressel2,qsedressel3,qsedressel4,qsedressel5,qsedressel6,lowthight,thexpans,PhysRevB.74.205318,lowthemp,qsedressel7,sonlarprb,PhysRevB.80.155404,matsci,PhysRevA.87.033603,aydin,sevan,nanotermo,top1,top2,PhysRevB.95.245432}. Examination of size dependent oscillations also has crucial importance in superconductors and topological insulators \cite{top1,top2,supcon1,supcon2,supcon3,PhysRevLett.109.226406}. Size and density dependent oscillations in aforementioned quantities are attributed to the quantization of energy spectrum and fluctuations in density of states around Fermi level at nanoscale \cite{qsedressel1,qsedressel2,qsedressel3,qsedressel4,qsedressel5,qsedressel6,Pal_2017}. For Fermi systems with linear energy dispersion relation (e.g. a Fermi system under an applied magnetic field or harmonic trap potential), quantum oscillations and its implications are extensively studied especially for specific materials \cite{PhysRevB.44.1646,qsedressel1,qsedressel2,qsedressel3,qsedressel4,qsedressel5,qsedressel6,ferry,revvmod}. For the systems obeying quadratic energy-momentum dispersion relation, on the other hand, there is a need for a mathematical model predicting quantum oscillations and explaining their nature in a comprehensive and material/case independent way. Moreover, a phase diagram separating oscillatory and stationary regimes on confinement-degeneracy space has never been proposed for any dispersion relation. Constructing a mathematical model and a phase diagram may help to deeply understand and efficiently control the oscillations to design novel nanoscale devices \cite{qsedressel6,PhysRevB.80.155404}.

In this article, we propose a theoretical model, half-vicinity model (HVM), that accurately predicts the oscillations appearing in confined Fermi gases. A rectangular confinement domain is chosen, because it is one of the most suitable geometries for common manufacturing methods of low dimensional nanostructures. In the following section, by examining the nature of TOV of quantum states, we introduce the model by defining a half-vicinity shell around Fermi level. $d$-dimensional discrete and continuum expressions of TOV function are given and thickness of half-vicinity shell, which plays an important role in HVM, is obtained for various dimensions. For 1D case, we obtain analytical expressions for TOV oscillations and their periods. In Sec. III, a phase diagram of quantum oscillations is established and phase transition interfaces between stationary (classical, continuous) and oscillatory (quantum, discrete) regimes are analytically given for 1D, 2D and 3D cases. Critical confinement and degeneracy values, which separates entirely stationary and oscillatory regions, are also examined by considering their aspect ratio dependencies. In Sec. IV, results of exact (definitional expressions based on infinite sums) and HVMs are compared for size and density dependent oscillations in the electronic specific heat capacity of strongly degenerate and confined ideal Fermi gases. Furthermore, broken equivalence of entropy-heat capacity in quantum degenerate limit of ideal Fermi gases is also well predicted by HVM. Finally, we discuss (Sec. V) the relationship between TOV and conventional density of states functions at Fermi level.

%%%        %%%%%%%%%%%%%%%%%%%%%%%%%%%%%%%%%%%%%%%%%%%%%%%%%%%%%%%%%%%        %%%
%%%        %%%%%%%%%%%%%%%%%%%%%%%% SECTION 2 %%%%%%%%%%%%%%%%%%%%%%%%        %%%
%%%        %%%%%%%%%%%%%%%%%%%%%%%%%%%%%%%%%%%%%%%%%%%%%%%%%%%%%%%%%%%        %%%

\section{Half-vicinity model for total occupancy variance}
Oscillations in thermodynamic and transport quantities originate from the nature of occupancy variance function. For this reason, in order to establish a model to understand the nature of oscillations, we need to examine TOV in detail.
\subsection{Total occupancy variance revisited}
Occupancies of quantum states for Fermions are described by Fermi-Dirac distribution function, $f=g_s/\left[\exp(-\Lambda+\tilde{\varepsilon})+1\right]$ where $g_s$ is spin degree of freedom, $\Lambda=\mu/(k_BT)$ is dimensionless chemical potential (or degeneracy parameter) indicating the strength of degeneracy, in which $\mu$ is chemical potential, $k_B$ is Boltzmann's constant, $T$ is temperature. $\tilde{\varepsilon}$ represents dimensionless energy eigenvalues (normalized to thermal energy $k_B T$). For quadratic dispersion relation of massive particles confined in a $d$-dimensional rectangular domain, $\tilde{\varepsilon}=\varepsilon/(k_BT)=(\alpha_1 i_1)^2+\cdots+(\alpha_d i_d)^2$ where $i_n$ is quantum state variable for a particular direction $n=\{1,2,\cdots,d\}$. Here $\alpha_n$ is confinement parameter indicating the strength of quantum confinement in a particular direction $n$ and defined as $\alpha_n=h/\left(\sqrt{8m k_B T}L_n\right)$ where $h$ is Planck's constant, $m$ is mass of the particle and $L_n$ is domain size in direction $n$.

Derivative of Fermi-Dirac distribution function with respect to $\Lambda$, a bell-shaped function peaked at Fermi level, is called occupancy variance function. Occupancy variance of an energy state $\tilde{\varepsilon}=\tilde{\varepsilon}(i_1,\ldots,i_d)$ is
\begin{equation}
\sigma^2(i_1,\ldots,i_d)=\frac{\partial f}{\partial \Lambda}=-\frac{\partial f}{\partial \tilde{\varepsilon}}=\frac{g_s}{4}\sech^2\left(\frac{\tilde{\varepsilon}-\Lambda}{2}\right).
\end{equation}
In literature, it's sometimes called also as thermal broadening function, since it represents thermal broadening of energy states around Fermi level \cite{datta}. Due to the fact that Fermi-Dirac distribution is a Bernoulli-type distribution (basically a sigmoid function), occupancy variance can also be represented by a Bernoulli-type variance form $\sigma^2=g_s\tilde{f}(1-\tilde{f})$, where $\tilde{f}=f/g_s$ is the spin normalized Fermi-Dirac distribution function. Spin factor $g_s$ is taken simply two in all numerical calculations in this article.

In order to understand the behavior of oscillatory quantities and how occupancy variance behaves under the accumulation operators such as summation or integration, we need to investigate the nature of TOV. TOV of Fermi particles inside a $d$-dimensional box is written in its exact form based on infinite sums as
\begin{equation}
\Sigma_D^2=\sum_{i_1=1}^{\infty}\cdots\sum_{i_d=1}^{\infty}\sigma^2(i_1,\ldots,i_d),
\end{equation}
where subscript $D$ stands for discrete calculations. When the summations in Eq. (2) are calculated by using the first two terms of Poisson summation formula (PSF) \cite{specfin,aydin}, the following expression can be obtained after some mathematical operations,
\begin{equation}
\begin{split}
\Sigma_{W}^2=
& g_s\sum_{n=0}^d\frac{(-1)^{d-n+1}\mkern+4mu\pi^{n/2}}{2^d \mkern+4mu d^{1-\Theta(n)\Theta(d-n)}}Li_{\frac{n-2}{2}}\left[-\exp(\Lambda)\right] \\
& \times\sum_{k=1}^{d}\prod_{m=k}^{k+n-1}{\alpha_{\text{mod}(m,d)+1}^{-1}},
\end{split}
\end{equation}
where $Li$ is polylogarithm function \cite{polylog} and $\Theta$ is left-continuous Heaviside step function. Since the expressions calculated by using the first two terms of PSF are equivalent to those obtained from Weyl conjecture \cite{specfin,pathbook,ddos}, we denote this expression by subscript $W$. In strongly degenerate conditions ($\Lambda>>1$), by using the asymptotic forms of polylogarithm functions \cite{polyexpbook}, Eq. (3) can be approximated by
\begin{equation}
\begin{split}
\Sigma_{W\mkern-5mu A}^2=
& g_s\sum_{n=0}^d\frac{(-1)^{d-n}\mkern+4mu\pi^{n/2}}{2^d \mkern+4mu d^{1-\Theta(n)\Theta(d-n)}}\frac{\Lambda^{(n-2)/2}}{\Gamma(n/2)} \\
& \times\sum_{k=1}^{d}\prod_{m=k}^{k+n-1}{\alpha_{\text{mod}(m,d)+1}^{-1}},
\end{split}
\end{equation}
where $\Gamma$ is gamma function.

When confinement parameters are much smaller than unity, namely in macroscale, summations may be converted directly into integrals by using continuum approximation and continuous TOV is obtained
\begin{equation}
\Sigma_C^2=-\frac{g_s\pi^{d/2}}{2^d \alpha_1 \cdots \alpha_d}Li_{\frac{d-2}{2}}\left[-\exp(\Lambda)\right],
\end{equation}
where $C$ subscript denotes the continuous expressions obtained under continuum approximation. In strongly degenerate case ($\Lambda>>1$), Eq. (5) can be approximated by using asymptotic expansions of polylogarithm functions
\begin{equation}
\Sigma_{C\mkern-5mu A}^2=\frac{g_s\pi^{d/2}}{2^d \alpha_1 \cdots \alpha_d}\frac{\Lambda^{(d-2)/2}}{\Gamma(d/2)}.
\end{equation}
As it should be, Eqs. (5) and (6) are actually the first (bulk) terms of Eqs. (3) and (4) respectively.

Expressions of thermodynamic and transport properties exhibiting oscillatory behavior contain Eq. (1) multiplied by some relevant functions of the quantity to be calculated. Then, it's proper to construct our HVM on TOV, in order to predict oscillatory behaviors in thermodynamic and transport properties.

\subsection{Basic concept of half-vicinity model}
Contributions to oscillatory physical quantities substantially come from partially occupied states around Fermi level. Under strongly confined and degenerate conditions, sharpness of the occupancy variance function around Fermi level increases and only few states contribute to TOV. In that case, it becomes possible to make a quantitative treatment of these states by constructing a shell which contains them inside. Therefore, for the accurate calculation of oscillatory physical quantities, instead of summing over all possible quantum states from unity to infinity, considering only those several states around Fermi level (inside the half-vicinity shell) is enough to represent the oscillations.

Half-vicinity shell in state space is defined as the interval containing the states in the half-vicinities ($\pm 1/2$) of Fermi level. It can be extended to any dimension, by considering the $\pm 1/2$ intervals of Fermi level in each direction at quantum state space. Magnitude of the contribution of each state to TOV is determined by its proximity to the Fermi level. Contribution of a state is maximum when it lies exactly on the Fermi level, and exponentially decays as the state is away from it. The number of half-vicinity states and their proximities to Fermi level change with degeneracy and/or confinement. When confinement becomes stronger, spacing between energy levels increases and causes a decrement in number of half-vicinity states. Because of this decrement, addition (removal) of a state to (from) half-vicinity shell leads to a strong change in TOV. This causes strong fluctuations both in number of half-vicinity states and their proximities to the Fermi level. In other words, half-vicinity states reorganize their distribution and proximities to the Fermi level when for instance sizes of the domain or density of the gas change. Each state gives distinct contribution to TOV. For macro domains with highly populated Fermi level, effect of this reorganization becomes statistically insignificant. Conversely, in confined and degenerate Fermi gases these reorganization of half-vicinity states manifest themselves as quantum oscillations in several thermodynamic and transport quantities due to changes in size and density.

For strongly confined and degenerate Fermi gases, TOV, Eq. (2), can approximately be calculated by restricting the summations over infinite number of quantum states to only the states inside the constructed half-vicinity shell. TOV based on HVM is then given by
\begin{equation}
\Sigma_{HV}^2=\sum_{i_1=1}^{\infty}\cdots\sum_{i_d=1}^{\infty}\sigma^2w_{HV}.
\end{equation}
Here, half-vicinity window function is define as
\begin{equation}
w_{HV}=\Theta\left(\Lambda-\tilde{\varepsilon}_{-1/2}\right)\Theta\left(\tilde{\varepsilon}_{1/2}-\Lambda\right),
\end{equation}
where $\Theta$ is Heaviside step function and $\tilde{\varepsilon}_{\pm1/2}=\alpha_1^2\left(i_1\pm 1/2\right)^2+\ldots+\alpha_d^2(i_d\pm 1/2)^2$ give upper and lower energy bounds of half-vicinity window. Window function $w_{HV}$ filters the states which do not significantly contribute to the oscillatory quantities and serves as a half-vicinity states pass filter, that restricts the upper and lower limits of summations. By definition, in state space, half-vicinity shell thickness is equal to unity in any particular direction, which is the minimum possible thickness in state space for each direction. On the other hand, in energy space its thickness varies with the strength of confinement and degeneracy. Half-vicinity window can equivalently be written as
\begin{equation}
w_{HV}=\Theta(\tilde{\varepsilon}_H-\tilde{\varepsilon})\Theta(\tilde{\varepsilon}-\tilde{\varepsilon}_L),
\end{equation}
where $\tilde{\varepsilon}_H=\Lambda+\delta_{\tilde{\varepsilon}_H}$, $\tilde{\varepsilon}_L=\Lambda-\delta_{\tilde{\varepsilon}_L}$, $\delta_{\tilde{\varepsilon}_H}=\delta_{\tilde{\varepsilon}}/2+\tilde{\varepsilon}_0/4$, $\delta_{\tilde{\varepsilon}_L}=\delta_{\tilde{\varepsilon}}/2-\tilde{\varepsilon}_0/4$. Here, the thickness of half-vicinity shell, $\delta_{\tilde{\varepsilon}}$, in its discrete form and the ground state energy in normalized energy space, $\tilde{\varepsilon}_0$, are given respectively as follows
\begin{subequations}
\begin{align}
& \delta_{\tilde{\varepsilon}}=2\left(\alpha_1^2 i_1+\ldots+\alpha_d^2 i_d\right) \\
& \tilde{\varepsilon}_0=\alpha_1^2+\ldots+\alpha_d^2
\end{align}
\end{subequations}
Instead of discrete definition of $\delta_{\tilde{\varepsilon}}$, it's also possible to define it based on continuous expressions which simplifies mathematical operations and will be given in the following subsection. Half-vicinity shell thickness is a crucial parameter since contributions to oscillatory properties mainly come from the states within this half-vicinity shell.

\begin{figure}[t]
\centering
\includegraphics[width=0.50\textwidth]{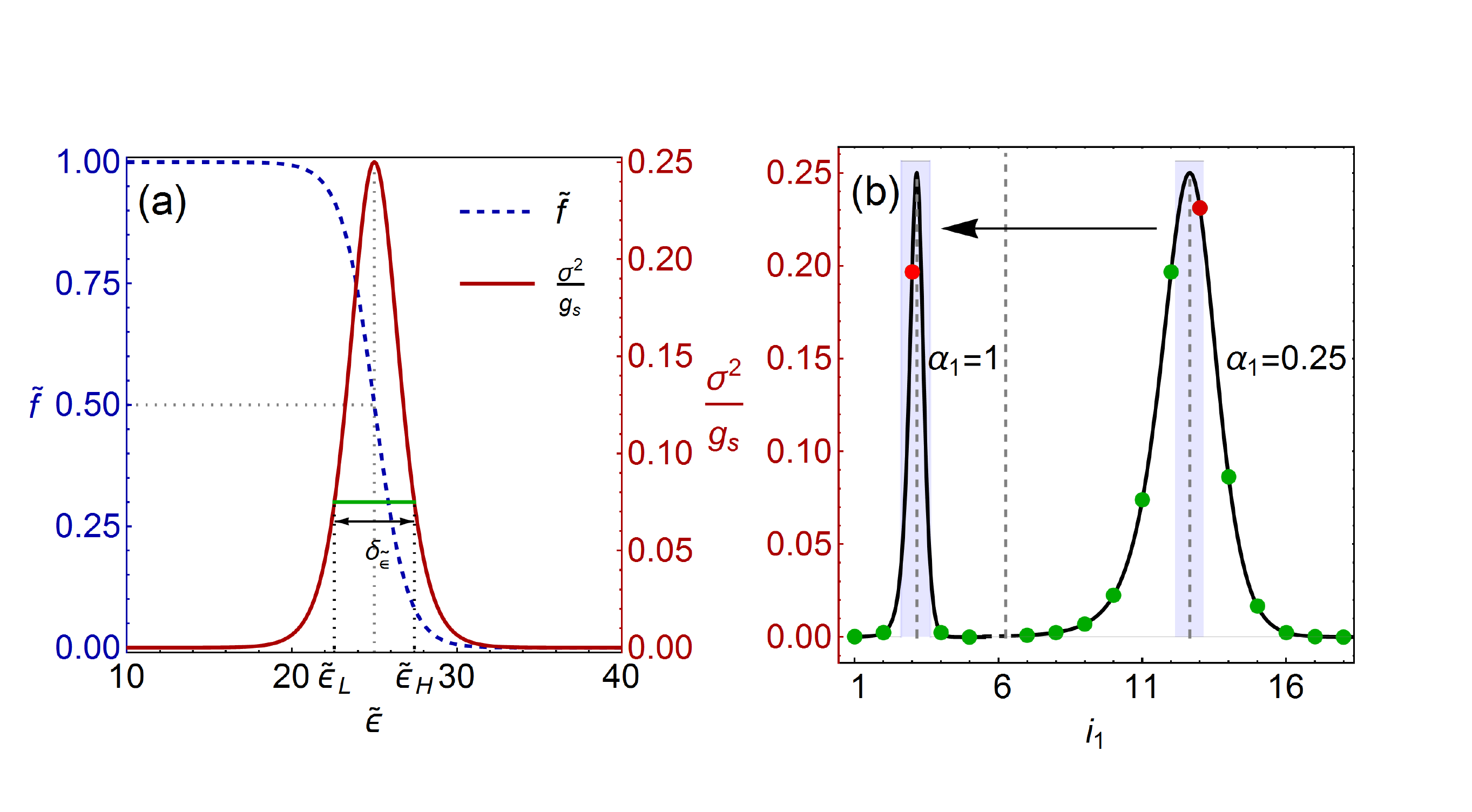}
\caption{(Color online) (a) Spin normalized 1D Fermi distribution and occupancy variance functions in energy space for $\Lambda=25$ and $\alpha=0.5$. Thickness of half-vicinity shell in energy space ($\delta_{\tilde{\varepsilon}}=5$) denoted by green line. (b) Spin normalized occupancy variance function in 1D state space for weakly confined ($\alpha_1=0.25$) and confined ($\alpha_1=1$) domains for a constant chemical potential ($\Lambda=10$). Shaded region denotes the thickness of half-vicinity shell which is equal to unity in state space. Red and green points represent the half-vicinity states (Fermi points in 1D) and off-half-vicinity states respectively.}
\label{fig:pic2}
\end{figure}

In Fig. 1(a), spin normalized 1D Fermi distribution and occupancy variance functions in energy space are shown by dashed-blue and solid-red curves respectively. Thickness of half-vicinity shell is represented by the solid horizontal green line. In Fig. 1(b), occupancy variance is plotted in 1D state space for two different confinement values for a constant chemical potential ($\Lambda=10$). As is seen from Fig. 1(b), when confinement increases (denoted by the leftward arrow), occupancy variance peak becomes sharper. Red states are the closest states to the respective Fermi levels, so their contributions are the largest. Light blue shaded region represents HV window. These closest states to Fermi level in 1D state space are called Fermi points. They constitute a Fermi line in 2D and Fermi surface in 3D cases. Green points represent the states outside of half-vicinity shell, which we call off-half-vicinity (OHV) states. Note that in strongly confined case contributions of OHV states are negligible. Thus, HVM represents the true behavior of occupancy variance function accurately in strongly degenerate cases, where oscillations are strong. On the contrary, for weakly confined or unconfined cases, where oscillations weaken or disappear, HVM representation of variance function becomes inaccurate because of non-negligible contributions of OHV states (Fig. 1(b) for $\alpha=0.25$). However, in this case, continuum expressions of variance function already represent the true behavior properly.

\subsection{Thickness of half-vicinity shell in energy space and renormalized HVM}
Thickness of half-vicinity shell in energy space plays an important role in HVM. To describe the thickness in a general way, we invoke number of states (NOS) and density of states (DOS) concepts. NOS inside the half-vicinity shell is found by $NOS_{HV}=\sum_{i_1=1}^{\infty}\cdots\sum_{i_d=1}^{\infty}w_{HV}$, which is an exact equation but does not lead to analytical expressions, which we seek here. Therefore, we use continuum approximation to convert summations into integrals and find the thickness of half-vicinity shell in energy space analytically as $\delta_{\tilde{\varepsilon}}=CNOS_{HV}/CDOS(\Lambda)$, where $CNOS_{HV}=\int_{0}^{\infty}\cdots\int_0^{\infty}w_{HV}di_1\ldots di_d$. By calculating the integrals of CNOS and using the known expressions of CDOS \cite{ddos}, $\delta_{\tilde{\varepsilon}}$ can be obtained for various dimensions in degenerate limit as given in Table I. Note that when $\sqrt{\Lambda}/\alpha_n$ value of a system becomes less than unity for a particular direction $n$, momentum modes in that direction becomes empty except the ground state and the system can be considered as if it has one dimension less.

\begin{table}[H]
\centering
\caption{CNOS, CDOS and thickness of half-vicinity shell in energy space for various dimensions.}
\label{my-label}
\def\arraystretch{1.75}
\setlength{\tabcolsep}{0.5em}
\begin{tabular}{ccccc}
\hline
Quantity 	                         & 1D & 2D & 3D  \\ \hline
$\frac{CNOS_{HV}}{g_s}$                   & $1$  & $\sqrt{\Lambda}\left(\frac{\alpha_1+\alpha_2}{\alpha_1 \alpha_2}\right)$ & $\frac{\pi\Lambda}{4}\left(\frac{\alpha_1+\alpha_2+\alpha_3}{\alpha_1 \alpha_2 \alpha_3}\right)$ \\
$\frac{CDOS(\Lambda)}{g_s}$                   & $\frac{1}{2\alpha_1\sqrt{\Lambda}}$  & $\frac{\pi}{4\alpha_1\alpha_2}$ & $\frac{\pi\sqrt{\Lambda}}{4\alpha_1\alpha_2\alpha_3}$ \\
$\delta_{\tilde{\varepsilon}}$			 & $2\alpha_1\sqrt{\Lambda}$ & $\frac{4\sqrt{\Lambda}}{\pi}(\alpha_1+\alpha_2)$ & $\sqrt{\Lambda}\left(\alpha_1+\alpha_2+\alpha_3\right)$  \\
\hline
\end{tabular}
\end{table}

$\delta_{\tilde{\varepsilon}}$ can be generalized into $d$-dimensions as
\begin{equation}
\delta_{\tilde{\varepsilon}}(d)=2\sqrt{\frac{\Lambda}{\pi}}\frac{\Gamma(d/2)}{\Gamma[(d+1)/2]}\sum_{n=1}^{d}\alpha_n.
\end{equation}

In continuum case, $(\alpha_1,\ldots,\alpha_d\rightarrow 0)$, HVM gives the following expression for TOV,
\begin{subequations}
\begin{align}
\Sigma_{HVC}^2 &=\int_0^{\infty}\sigma^2 w_{HV}di_1\ldots di_d \\
&=\int_0^{\infty}\sigma^2 w_{HV}CDOS(\tilde{\varepsilon})d\tilde{\varepsilon} \\
&=\int_{\tilde{\varepsilon}_L}^{\tilde{\varepsilon}_H}\sigma^2 CDOS(\tilde{\varepsilon})d\tilde{\varepsilon}.
\end{align}
\end{subequations}
It should be noted that $\tilde{\varepsilon}_0\rightarrow 0$ in continuous case and $\tilde{\varepsilon}_H$ and $\tilde{\varepsilon}_L$ become $\Lambda+\delta_{\tilde{\varepsilon}}/2$ and $\Lambda-\delta_{\tilde{\varepsilon}}/2$ respectively. It's possible to obtain an analytical solution for the integral in Eq. (12c) for any dimension by considering strongly degenerate conditions $(\Lambda>>1)$,
\begin{equation}
\Sigma_{HVC}^2\cong\tanh\left(\frac{\delta_{\tilde{\varepsilon}}}{4}\right)\Sigma_{CA}^2.
\end{equation}
Here, $\tanh$ factor acts as an amplitude renormalization factor. Although this factor is found by a continuous approach, the similar renormalization factor is expected also for the discrete case, because amplitude renormalization process is independent of the type of accumulation operators which is verified by the numerical results in this article. Thus, we may do the following approximation
\begin{equation}
\begin{aligned}
\Sigma_{HV}^2 &=\sum_{i_1=1}^{\infty}\cdots\sum_{i_d=1}^{\infty}\sigma^2w_{HV} \\
&\cong\tanh\left(\frac{\delta_{\tilde{\varepsilon}}}{4}\right)\left[\sum_{i_1=1}^{\infty}\cdots\sum_{i_d=1}^{\infty}\sigma^2\right]=\tanh\left(\frac{\delta_{\tilde{\varepsilon}}}{4}\right)\Sigma_{D}^2.
\end{aligned}
\end{equation}
Then, we can express discrete TOV, $\Sigma_{D}^2$, in terms of $\Sigma_{HV}^2$ as follows,
\begin{equation}
\Sigma_{D}^2\approx\frac{\Sigma_{HV}^2}{\tanh\left(\delta_{\tilde{\varepsilon}}/4\right)}=\Sigma_{RHV}^2,
\end{equation}
where $RHV$ subscript denotes renormalized half-vicinity. Factor of $1/\tanh$ renormalizes the amplitude of variance function found by HVM (which only considers HV states) in order to represent also the contributions of OHV states. In other words, $\Sigma_{RHV}^2$ is the renormalized TOV based on HVM and it includes contributions of both HV and OHV states. Accuracy of this expression depends on the thickness of half-vicinity shell inside the factor of $\tanh$, which is a function of confinement and degeneracy. The remarkable accuracy of Eq. (15) in comparison with the exact equation (Eq. (2)) is shown in Fig. 2. Although Weyl representation (Eq. (3)) does not predict the oscillations but only their trend, HVM quite perfectly matches with the exact results especially for high degeneracy and confinement conditions where oscillations are significant. Consequently, by considering Eqs. (2), (7) and (15), HVM proposes the following transformation:
\begin{equation}
\sigma^2\xrightarrow{HVM}\sigma^2\frac{w_{HV}}{\tanh\left(\delta_{\tilde{\varepsilon}}/4\right)}
\end{equation}
for the calculation of any thermodynamic and transport quantity containing $\sigma^2$ term. By this transformation, infinite sums are converted into finite sums over the minimum number of states near to Fermi level. As is clear from the comparisons in Fig. 2, HVM can be used as a reliable tool to predict and represent quantum oscillations.

\begin{figure*}[t]
\centering
\includegraphics[width=0.99\textwidth]{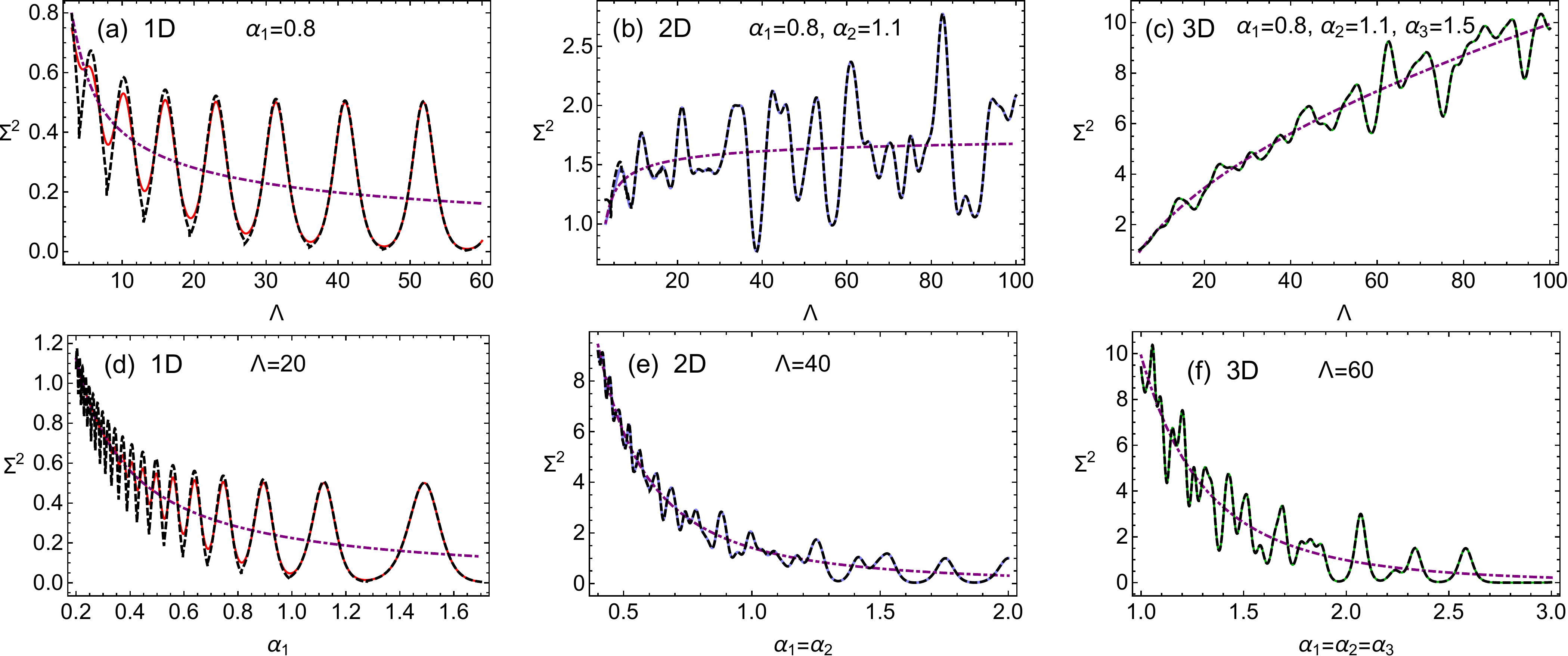}
\caption{(Color online) Comparison of TOV results based on exact model, HVM and Weyl expressions. (a) 1D (b) 2D (c) 3D occupancy variance functions changing with degeneracy and (d) 1D (e) 2D (f) 3D occupancy variance functions changing with isometric confinement in all available directions. Red, blue and green solid curves represent Eq. (2) in 1D, 2D and 3D respectively. Eq. (15) (HVM estimation) is represented by dashed black curves. Dot-dashed purple curves are Weyl representations of TOV functions from Eq. (3).}
\label{fig:pic4}
\end{figure*}

\subsection{Analytical expression for TOV in 1D case}
In highly degenerate and strongly confined 1D case, summation of TOV in Eq. (2) vanishes and HVM represents only the contribution of Fermi point inside $i_F\pm1/2$ interval in state space, where $i_F=\sqrt{\Lambda}/\alpha_1$. Therefore, for this special case, there is no need to do summation over half-vicinity shell states (since there is only one) and a special value of that state variable can be found analytically as
\begin{equation}
i_{*}=\left[i_F\right]=i_F-\frac{1}{\pi}\arctan{\left[\tan\left(\pi i_F\right)\right]},
\end{equation}
where $\left[\cdots\right]$ denotes the round function which can be represented by elementary functions on the right hand side. Then, renormalized TOV in 1D case is
\begin{equation}
\Sigma_{RHV}^2=\frac{\sigma^2(i_{*})}{\tanh\left(\alpha_1\sqrt{\Lambda}/2\right)}.
\end{equation}

A comparison for variations of exact and HVM's TOV functions with confinement is given in Fig. 3. Solid red, solid blue and dashed black curves are results of Eqs. (2), (5) and (18) respectively. Accuracy of HVM in representing the prominent oscillations is quite good, even though it only considers contribution of Fermi points. Since Eqs. (3) and (5) are equal to each other for 1D case, there is no need to examine $\Sigma_W^2$ additionally. It's seen from the red curve in Fig. 3 that transitions between stationary and oscillatory regimes are smooth and $\ln 3/\sqrt{\Lambda}$ line perfectly separates these regimes, which will be examined in detail in the following section.

\begin{figure}[H]
\centering
\includegraphics[width=0.45\textwidth]{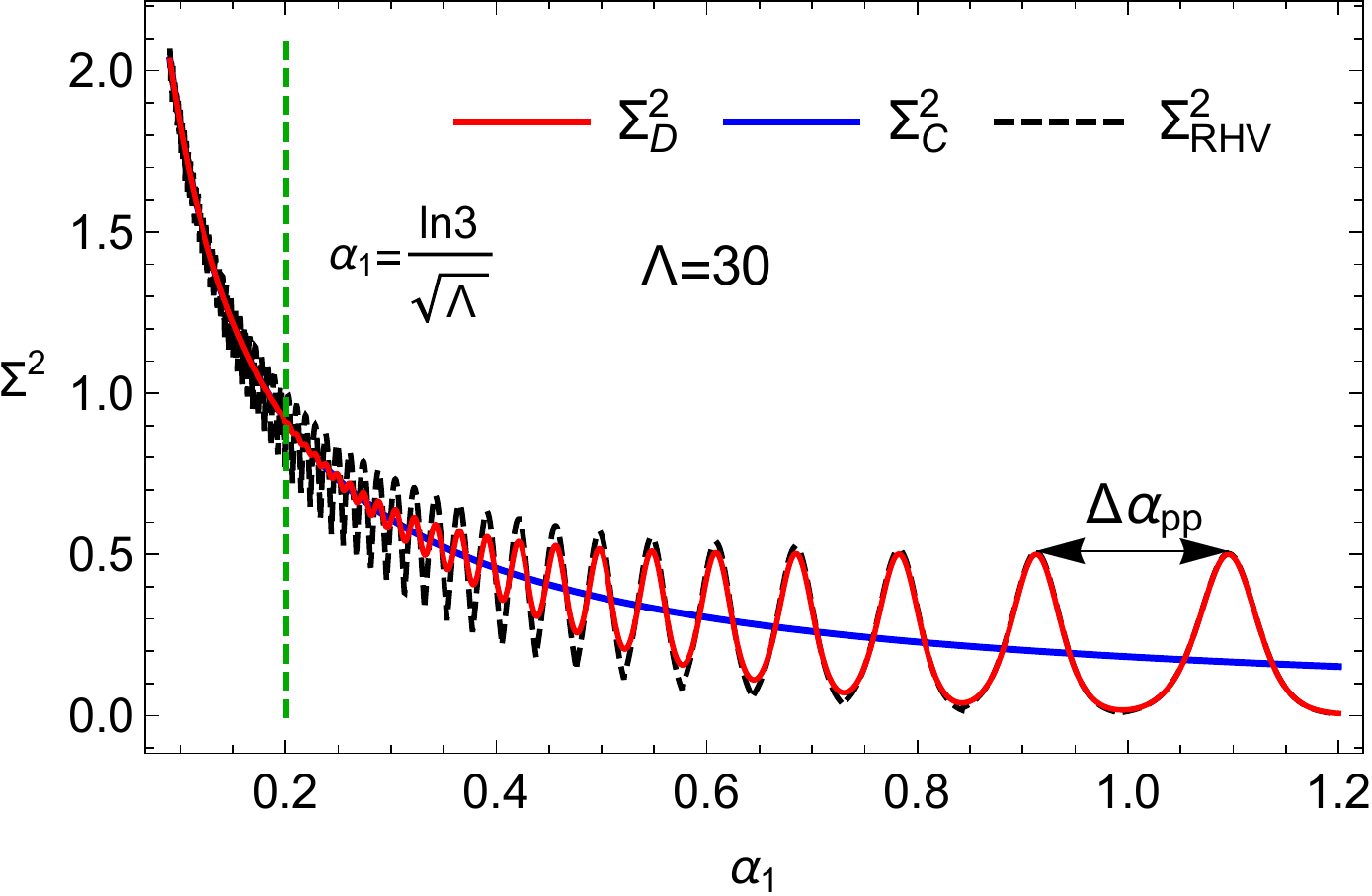}
\caption{(Color online) A comparison of exact and HVM results for the variation of TOV with confinement when $\Lambda=30$. Green vertical line, $\alpha_1=\ln 3/\sqrt{\Lambda}$, separates the oscillatory regime from the stationary one. $\Delta\alpha_{pp}$ denotes the period of TOV varying with confinement.}
\label{fig:pic3}
\end{figure}

Extremum points of oscillations in Fig. 3 correspond to the values where the total number of particles is integer. We can determine the periodicity of TOV changing with $\Lambda$ and $\alpha$, considering the fact that maxima and minima of oscillations correspond to odd and even integer particle numbers respectively, for $g_s=2$. Special to the 1D case, expression found by considering first two terms of PSF exactly represent number of particles for a given chemical potential in degenerate case \cite{aydin}. We find the values of $\Lambda$ and $\alpha$ correspond to integer numbers of particles as $\Lambda=[\alpha_1(\tilde{N}+1/2)]^2$ and $\alpha_1=\sqrt{\Lambda}/(\tilde{N}+1/2)$ respectively. Periods of oscillations in $\alpha$ and $\Lambda$ spaces can then easily be found respectively as
\begin{subequations}
\begin{align}
& \Delta\alpha_{pp}=\frac{4\sqrt{\Lambda}}{(2\tilde{N}+1)(2\tilde{N}+3)}=\frac{2\alpha_1}{2\tilde{N}+3}, \\
& \Delta\Lambda_{pp}=2\alpha_1^2(\tilde{N}+1)=8\Lambda\frac{\tilde{N}+1}{(2\tilde{N}+1)^2},
\end{align}
\end{subequations}
where $\tilde{N}=N/g_s$ is spin normalized particle number.

Considering the fact that $\Sigma_{RHV}^2$ mimics the oscillations in an almost perfect way, an intriguing question comes to the mind: Is it possible to determine analytically where quantum oscillations start as the confinement and/or degeneracy changes?

%%%        %%%%%%%%%%%%%%%%%%%%%%%%%%%%%%%%%%%%%%%%%%%%%%%%%%%%%%%%%%%        %%%
%%%        %%%%%%%%%%%%%%%%%%%%%%%% SECTION 3 %%%%%%%%%%%%%%%%%%%%%%%%        %%%
%%%        %%%%%%%%%%%%%%%%%%%%%%%%%%%%%%%%%%%%%%%%%%%%%%%%%%%%%%%%%%%        %%%

\section{A phase diagram for quantum oscillations: Transition from classical to quantum behavior}
HVM takes the discreteness of quantum states into account and allows us to accurately calculate the thermodynamic and transport properties exhibiting size and density dependent quantum oscillations without using infinite summations. In addition to that, HVM can accurately estimate where the transition from classical behavior to quantum behavior starts, in the framework of quantum oscillations.

The states contributing to TOV consist of HV states and OHV states (\textit{i.e.}, $\Sigma_{D}^2=\Sigma_{HV}^2+\Sigma_{OHV}^2$). HV states are responsible from the oscillatory part of TOV, whereas OHV states represent the stationary part. They constitute two competing parts of TOV. Domination of contributions of HV states over OHV ones leads to oscillations and vice versa. Hence, by considering $\Sigma_{HV}^2=\Sigma_{OHV}^2$ balance, we can compare contributions of HV states and OHV states to define a universal (material independent) recipe for the separation of stationary regime (SR) from oscillatory regime (OR). According to the recipe, when $\Sigma_{HV}^2<\Sigma_{OHV}^2$, oscillations disappear (SR) and on the opposite condition $\Sigma_{HV}^2>\Sigma_{OHV}^2$, oscillations reveal (OR). Since $\Sigma_{OHV}^2=\Sigma_{D}^2-\Sigma_{HV}^2$, SR-OR transition can be quantified as the balance of the contributions of HV and OHV states ($\Sigma_{HV}=\Sigma_{OHV}$) by $\Sigma_{D}^2=2\Sigma_{HV}^2$.

Since the transition is not sharp but smooth, we can safely use analytical expressions of TOV, instead of their exact expressions, to find an analytical expression for SR-OR separation. From the balance condition between contributions of HV and OHV states as well as Eq. (13), we can determine the following analytical condition for the transition between SR and OR,
\begin{equation}
\Sigma_{CA}^2=2\Sigma_{HVC}^2 \;\Rightarrow\; 1=2\tanh\left(\delta_{\tilde{\varepsilon}}/4\right) \;\Rightarrow\; \delta_{\tilde{\varepsilon}}=2\ln 3.
\end{equation}
It's seen from red curve in Fig. 3, balance condition quantified in Eq. (20) clearly gives the SR-OR separation. By decreasing confinement or degeneracy, the number of states around Fermi level increases, contributions of OHV states become also appreciable and their contributions make oscillation amplitude smaller. The more number of states around Fermi level, the smaller oscillation amplitude. When contributions of OHV states exceeds that of HV states, oscillations disappear.

To complete the construction of the phase diagram, it is necessary to check also the number of particles, ($N$), inside the system, since we are dealing with extremely confined systems having relatively low number of particles. For statistical representations there has to be sufficiently large number of particles inside the system. Nevertheless, the physically meaningful region in a phase diagram can be stated by the condition $N\geq 1$. Full list of recipes and conditions to establish the phase diagram is given in the Table 2.

\begin{table}[t]
\centering
\caption{Recipes and conditions for the construction of the phase diagram of quantum oscillations.}
\label{my-label}
\def\arraystretch{1.6}
\setlength{\tabcolsep}{1em}
\begin{tabular}{cccc}
\hline
Region 		 & Recipe & Condition  \\ \hline
OR 				 & $\Sigma_{HV}^2> \Sigma_{OHV}^2$ & $\delta_{\tilde{\varepsilon}}>2\ln 3$  \\
SR 			 & $\Sigma_{HV}^2<\Sigma_{OHV}^2$ & $\delta_{\tilde{\varepsilon}}<2\ln 3$  \\
Unphysical & $N<1$ & $\sum_{i_1,\ldots,i_d=1}^{\infty}f<1$ \\
\hline
\end{tabular}
\end{table}

According to the recipes and conditions given in Table 2, phase diagrams of quantum oscillations in degeneracy-confinement space for various dimensions can be determined. For isometric 3D domains, the phase diagram is constructed and given in Fig. 4. Solid-black curves represent interfaces defined by the conditions in Table II. $\delta_{\tilde{\varepsilon}}=2\ln 3$ condition, representing the balance of HV and OHV states, separates OR and SR. When both confinement and degeneracy sufficiently increase, system enters into the quantum regime where certain thermodynamic and transport quantities oscillate with varying size or density.

\begin{figure}[H]
\centering
\includegraphics[width=0.45\textwidth]{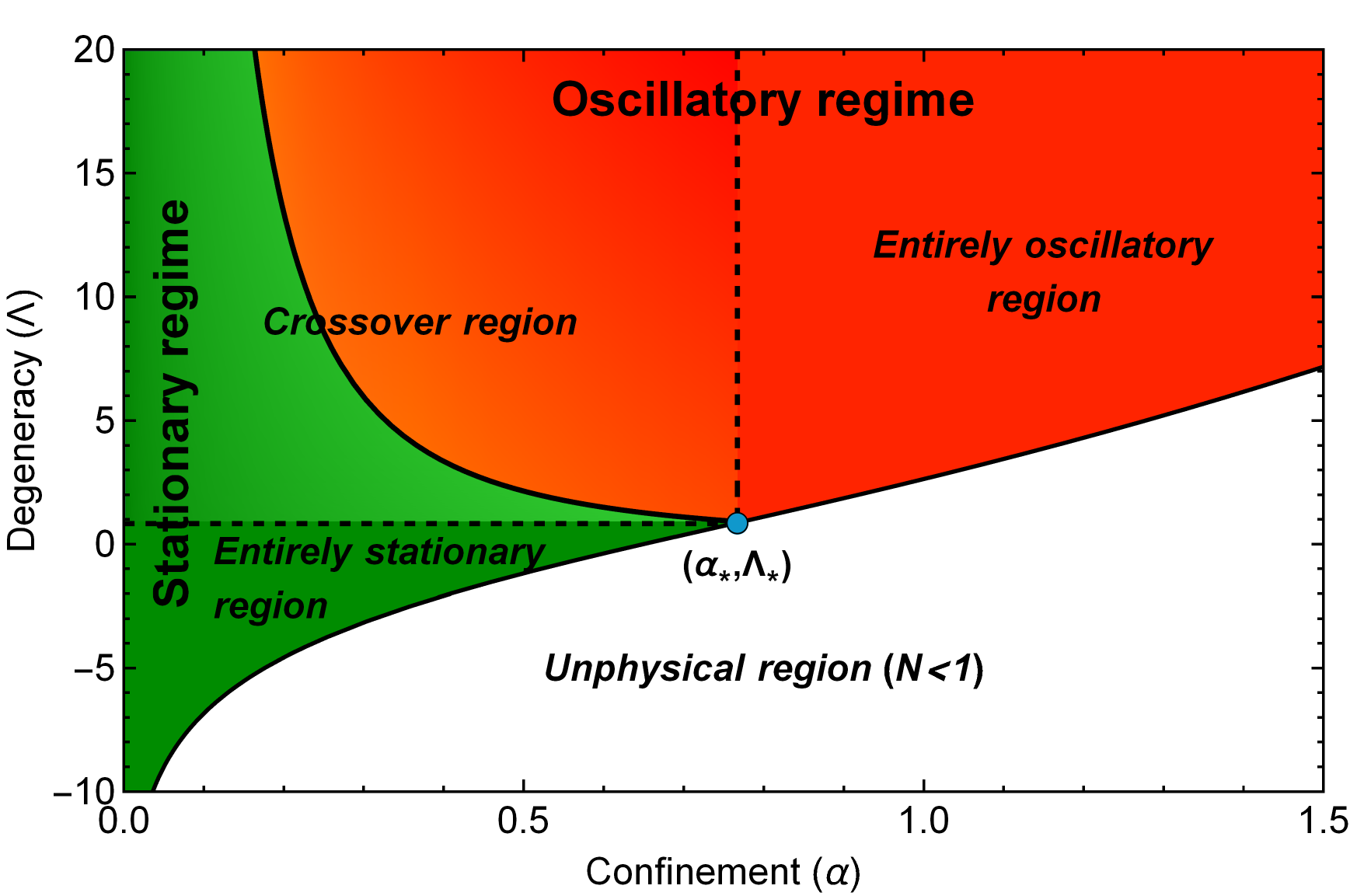}
\caption{(Color online) A phase diagram on degeneracy-confinement space for quantum oscillations. $\delta_{\tilde{\varepsilon}}=2\ln3$ condition defines the boundary between stationary (classical) and oscillatory (quantum) regimes. Blue dot represents the critical point in the phase diagram for isometric 3D domains.}
\label{fig:pic4}
\end{figure}

Intersection of SR-OR interface curve with $N=1$ curve denotes the critical point which is represented by blue dot in Fig. 4. Critical value of the confinement parameter at this point is $\alpha_{*}^{3D}=0.78$ and the corresponding degeneracy value is $\Lambda_{*}^{3D}=0.88$. These values are universal for isometric rectangular confinement domains. For confinement values below $\alpha_{*}$, existence of oscillations can be controlled and they can even be suppressed by decreasing degeneracy (through density or temperature). Similarly, for degeneracy values higher than $\Lambda_*$, existence of oscillations can be controlled by changing confinement. This region defined by $\{\Lambda>\Lambda_{*},\alpha<\alpha_*\}$ is the crossover region restricted by dashed lines in the phase diagram. On the contrary, for higher confinement values than $\alpha_{*}$, quantum oscillations cannot be suppressed and the region is called entirely oscillatory region. Below the critical degeneracy values, system does not exhibit oscillatory behaviors regardless of the values of confinements. Hence, this critical point is used to define different regions (crossover region, entirely OR and entirely SR). Regimes on phase diagram and corresponding conditions are summarized in Table III.

\begin{table}[H]
\centering
\caption{Conditions and regions of the phase diagram.}
\label{my-label}
\def\arraystretch{1.6}
\setlength{\tabcolsep}{1em}
\begin{tabular}{cccc}
\hline
Conditions & Regions  \\ \hline
$\alpha<\alpha_{*},\;\; \Lambda<\Lambda_*$ & Entirely SR  \\
$\alpha<\alpha_{*},\;\; \Lambda>\Lambda_*$ & Crossover region  \\
$\alpha>\alpha_{*},\;\; \Lambda>\Lambda_*$ & Entirely OR \\
$\alpha>\alpha_{*},\;\; \Lambda<\Lambda_*$ & Unphysical region \\
\hline
\end{tabular}
\end{table}

Although the phase diagram in Fig. 4 represents only isometric 3D domain, the form of the phase diagrams for 1D and isometric 2D domains are also very similar to the 3D one. Only SR-OR interface slightly shifts upward while $N=1$ curve slightly rotates in clockwise direction around the critical point as the dimension decreases. For 1D and isometric 2D domains, universal critical values can also be found as $\alpha_{*}^{1D}=1.07$, $\alpha_{*}^{2D}=0.87$ and $\Lambda_{*}^{1D}=1.05$, $\Lambda_{*}^{2D}=0.98$.

It's also possible to investigate the anisometric domains by defining aspect ratios. For 2D domain, $r_{12}=L_1/L_2=\alpha_2/\alpha_1$ denotes the aspect ratio. For 3D, additionally we define $r_{13}=L_1/L_3=\alpha_3/\alpha_1$. Here, the most confined direction is chosen as direction $1$. In Fig. 5, we show how the critical confinement values are changing with respect to aspect ratios of 2D and 3D domains.

\begin{figure}[H]
\centering
\includegraphics[width=0.47\textwidth]{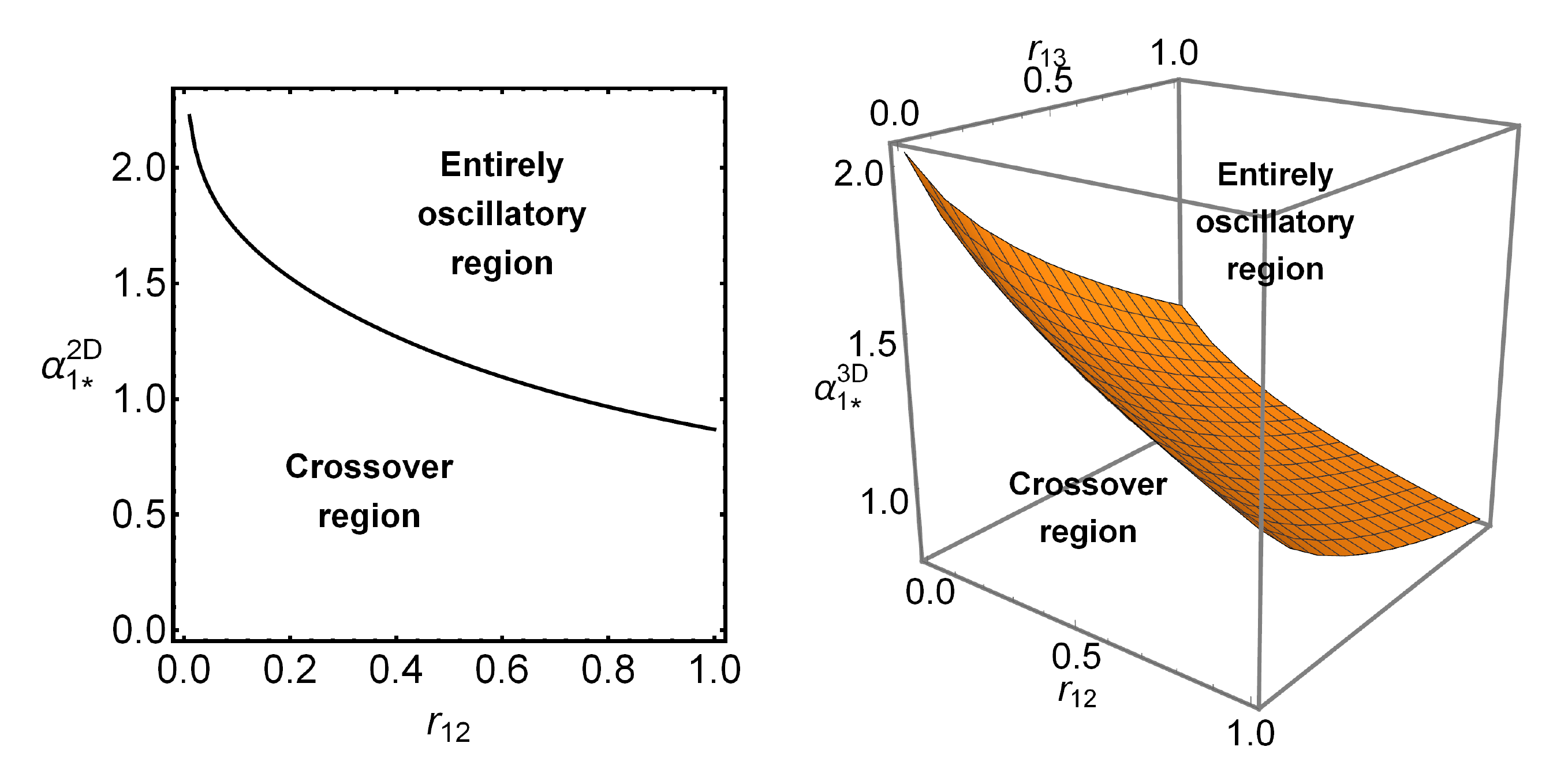}
\caption{(Color online) Aspect ratio dependencies of critical confinement values for 2D and 3D cases.}
\label{fig:pic4}
\end{figure}

Although the values in Fig. 5 are calculated in an exact way by using Tables I and II, $\alpha_{1*}^{2D}$ can be approximated by $\alpha_{1*}^{2D}\approx 0.87-0.40\ln r_{12}$ with less than $1.5\%$ mean absolute percentage error (MAPE) for $0.1\leq r_{12} \leq 1$ interval. Similarly, $\alpha_{1*}^{3D}$ is approximated by $\alpha_{1*}^{3D}\approx 0.78-0.27\ln r_{12}-0.27\ln r_{13}$ with less than $2.2\%$ MAPE for $0.1\leq \{r_{12},r_{13}\} \leq 1$ intervals.

Critical degeneracy values for anisometric cases can easily be found for 3D, 2D and 1D cases respectively,
\begin{subequations}
\begin{align}
&\Lambda_{*}^{3D}=\left[\frac{2\ln 3}{\alpha_{1*}^{3D}(1+r_{12}+r_{13})}\right]^2, \\
&\Lambda_{*}^{2D}=\left[\frac{\pi\ln 3}{2\alpha_{1*}^{2D}(1+r_{12})}\right]^2, \\
&\Lambda_{*}^{1D}=\left(\frac{\ln 3}{\alpha_{1*}^{1D}}\right)^2.
\end{align}
\end{subequations}

It's important to note that thickness of half-vicinity shell indicates the quantumness of the system. When volume $V\rightarrow\infty$ or temperature $T\rightarrow\infty$ or density $n\rightarrow 0$, thickness $\delta_{\tilde{\varepsilon}}\rightarrow 0 <2\ln3$ and system behaves classically. Conversely, for sets of $(V,T,n)$ making $\delta_{\tilde{\varepsilon}}>2\ln3$, the system shows quantum oscillatory behaviors. As a matter of fact, $\delta_{\tilde{\varepsilon}}$ is a precise indicator of whether system is in quantum regime or not in terms of oscillations.

In order to see the full picture of quantum oscillations in 1D case, variation of TOV, Eq. (2), with confinement and degeneracy is given in Fig. 6. It is seen now even more clearly that $\delta_{\tilde{\varepsilon}}=2\ln3$ plane perfectly separates oscillatory region from the stationary one regardless of the values of $\alpha$ and $\Lambda$. Increment in degeneracy and/or confinement make $\delta_{\tilde{\varepsilon}}$ larger and cause strong oscillations.

\begin{figure}[H]
\centering
\includegraphics[width=0.35\textwidth]{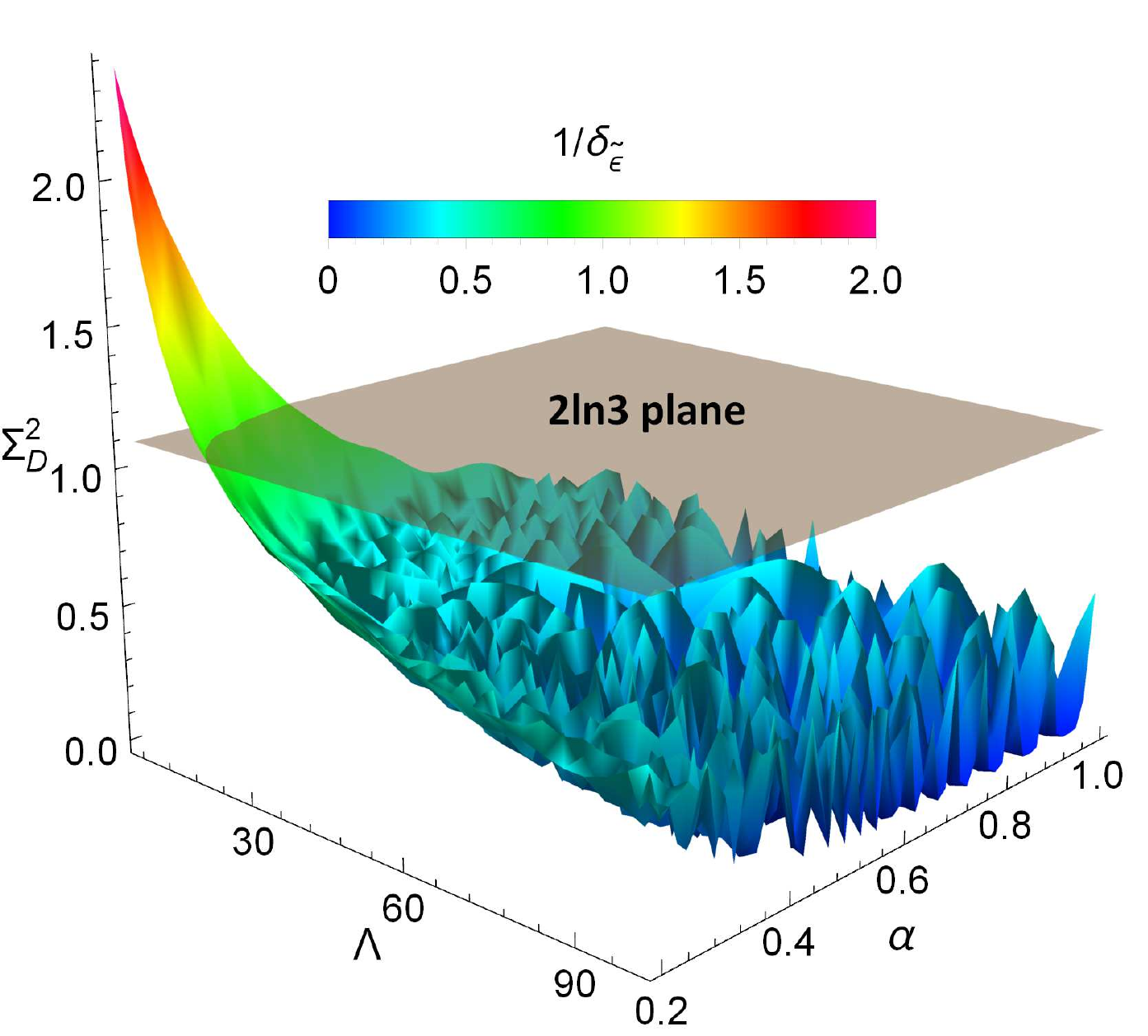}
\caption{(Color online) Variation of discrete TOV ($\Sigma^2_D$) with confinement ($\alpha$) and degeneracy ($\Lambda$). Gray $2\ln3$ plane separates the oscillatory (quantum) and stationary (classical) regimes.}
\label{fig:pic5}
\end{figure}

HVM is constructed to predict oscillations appearing in strongly confined and highly degenerate conditions. Accuracy of its predictions increases with increasing confinement and degeneracy, while accuracy becomes weaker for the regions near to SR-OR interface in phase diagram, Fig. 4. To be precise, HVM predicts oscillations almost perfectly, as long as confinement and degeneracy are larger than their critical values ($\alpha_*$ and $\Lambda_*$).

%%%        %%%%%%%%%%%%%%%%%%%%%%%%%%%%%%%%%%%%%%%%%%%%%%%%%%%%%%%%%%%        %%%
%%%%%%%%%%%%%%%%%%%%%%%%%%%%%%%%%%% SECTION 4 %%%%%%%%%%%%%%%%%%%%%%%%%%%%%%%%%%%
%%%        %%%%%%%%%%%%%%%%%%%%%%%%%%%%%%%%%%%%%%%%%%%%%%%%%%%%%%%%%%%        %%%

\section{Accuracy of HVM and phase diagram in predicting oscillations in thermodynamic quantities}
\subsection{Size and density dependent oscillations in electronic heat capacity}
In previous sections we investigated the oscillatory behavior of TOV function and constructed a theoretical model (HVM) for its calculation. In this subsection, we examine quantum oscillations of heat capacity of an electron gas confined in a nanowire by considering both exact and HVM. Electronic heat capacity from the derivative of internal energy with respect to temperature is written in its exact (based on infinite sums) form as
\begin{equation}
C_V=g_s k_B\left[\sum_{i_n=1}^{\infty}\tilde{\varepsilon}^2\sigma^2-\frac{\left(\sum_{i_n=1}^{\infty}\tilde{\varepsilon}\sigma^2\right)^2}{\sum_{i_n=1}^{\infty}\sigma^2}\right].
\end{equation}
Heat capacity expression, Eq. (22), contains zeroth, first and second order energy moments of occupancy variance. Each summation in Eq. (22) has distinct oscillatory behavior, which leads to the characteristic oscillations in heat capacity. In order to reveal quantum oscillations and make electronic contribution of heat capacity appreciable, we focus on very low temperature ($T=5$K), high electron density $(\sim 10^{25} \text{m}^{-3})$ and nanoscale confinements in which quantum effects are observable. To make our discussions quantity-independent, we examine electronic heat capacity per particle $(c_V=C_V/N)$. In the calculations, chemical potential is obtained numerically from particle number equation as functions of temperature, density and confinement.

In Fig. 7, accuracy of HVM is examined by considering size and density dependent oscillations of normalized electronic specific heat. $c_V^0$ represents the specific heat expression under continuum approximation, where all summations in Eq. (22) replaced by integrals. In all cases in Fig. 7, an electron gas confined in a quantum wire with 200 nm long $(L_3)$ is considered at 5K temperature. Red, blue and green curves in Fig. 7(a), 7(b) and 7(c) respectively represent the results of the exact model, while the black dashed curves are the results of HVM. Although confinement domain is a nanowire, it cannot be considered as a pure 1D structure since there are still a few excited states in transverse directions. Thus, calculations are done over triple sums.

\begin{figure*}[t]
\centering
\includegraphics[width=0.99\textwidth]{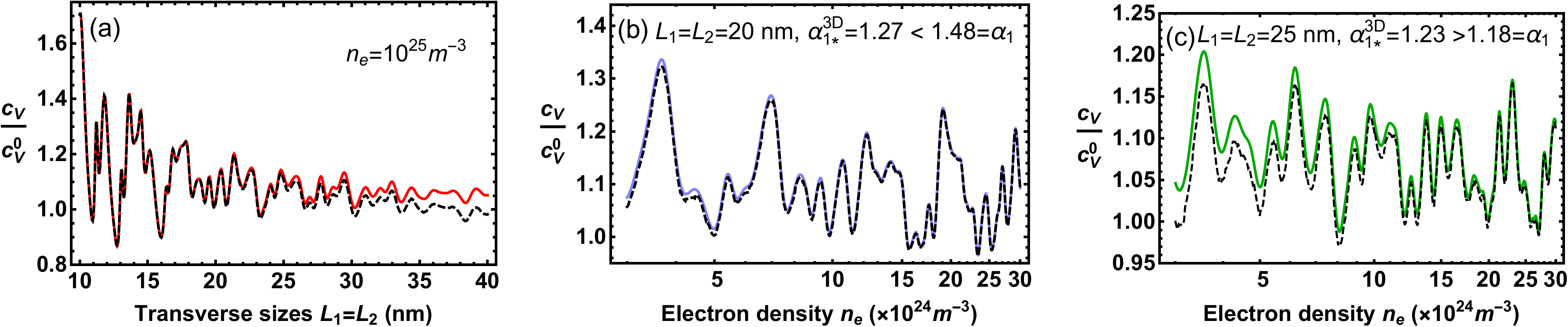}
\caption{(Color online) Accuracy of HVM on predicting (a) size dependent and (b), (c) density dependent oscillations in normalized heat capacity for a nanowire with 200 nm long at 5K temperature. Solid curves are the results obtained by using exact expressions (based on infinite sums), while dashed black curves are obtained by using HVM.}
\label{fig:pic7}
\end{figure*}

In Fig. 7(a), sizes of the domain in transverse directions ($L_1$ and $L_2$) are changed simultaneously from 10 nm to 40 nm for $n=10^{25}\text{m}^{-3}$ where $\Lambda$ is always larger than $\Lambda_{*}$. Confinement in transverse directions spans the values larger and smaller than $\alpha_*$. HVM predicts oscillatory size dependence of electronic heat capacity pretty well between 10 nm and 25 nm, where confinement is strong, oscillations are large and unavoidable ($\alpha\geq\alpha_*$). For 25 nm and 40 nm interval, however, confinement becomes weaker ($\alpha\leq\alpha_*$) and an appreciable difference appears in between exact and HVM results while oscillations are also weak. This is an expected result, since the system approaches to SR-OR interface in phase diagram.

In Fig. 7(b), quantum oscillations in heat capacity varying with electron density are shown for a square wire having 20 nm size in transverse directions. Note that density variation for a constant confinement corresponds to a vertical movement on the phase diagram. HVM perfectly matches with the exact model almost everywhere in Fig. 7(b) because of strong confinement, $\alpha_{1}=\alpha_{2}=1.48>\alpha_{1*}^{3D}=1.27$. System shows oscillatory behavior regardless of the value of degeneracy (or electron density). In other words, oscillations are unavoidable for the system discussed in Fig. 7(b) as long as the confinement is sustained. By increasing the domain size from 20 nm to 25 nm in confined directions (reducing the confinement), we can put the system into the crossover regime where $\alpha_{1}=\alpha_2=1.18<\alpha_{1*}^{3D}=1.23$ as it's shown in Fig. 7(c). For this case, HVM does not match well with the exact model for lower electron densities, where the system is getting closer to SR-OR interface. Oscillations decay by decreasing confinement and degeneracy but the decay rate of oscillations with degeneracy depends on how much the confinement smaller than the critical confinement value. Since the confinement is not much smaller than $\alpha_*$, there is no considerable decay in oscillations in Fig. 7(c). All these behaviors confirm the predictions of the phase diagram presented in Fig. 4 as well as aspect ratio dependencies of critical confinement parameters given in Fig. 5.

In contrast to size and density, variation of temperature does not lead to oscillatory behaviors in thermodynamic and transport quantities. The reason can clearly be seen from Eqs. (17) and (18) where oscillations are originated from the variation of round of $i_F$. Since temperature dependence of chemical potential disappears in strongly degenerate case, $[i_F]=[\sqrt{\Lambda}/\alpha]$ becomes independent of temperature. Hence, in strongly degenerate case, temperature variation does not affect the proximities of states to the Fermi level in state space, but just changes the occupation probabilities of the states. In other words, thermal broadening happens only in energy space, not in state space. Since the proximities do not variate, oscillations do not appear. This explanation can be directly extended to higher dimensional cases also, due to the orthogonality of state space dimensions.

\subsection{Oscillatory violation of entropy-heat capacity equivalence in degenerate limit}
Continuum expressions of entropy and heat capacity of an ideal Fermi gas are equal to each other in degenerate limit \cite{pathbook}. When we consider the discrete nature of quantum states, however, entropy-heat capacity equivalence is also broken and quantum behaviors lead to an oscillatory non-equivalence of entropy and heat capacity of an ideal degenerate Fermi gas at nanoscale.

The well-known form of entropy is written as
\begin{equation}
S=-g_s k_B\left[\sum_{i_n=1}^{\infty}{\tilde{f}\ln(\tilde{f})+(1-\tilde{f})\ln(1-\tilde{f})}\right].
\end{equation}
In degenerate limit, both Eq. (22) and (23) give the same expression (for instance, $g_s\pi^2Nk_B/(2\Lambda)$ for 3D case) under continuum approximation. However, as it is seen from Fig. 8, this equivalence is broken. Heat capacity/entropy ratio deviates from unity and variates with domain size and density in an oscillatory fashion. A nanowire in the previous subsection is reconsidered with the same specifications here for the examination of heat capacity-entropy ratio. Both electronic heat capacity and entropy oscillate in confined and degenerate conditions. Although their scales are almost the same, due to the phase and magnitude differences of both function's characteristic oscillations, their ratio also oscillates around unity. Increasing domain size or temperature decreases the amplitude of oscillations. Similar to heat capacity results, the success of HVM is quite good for higher electron density and confinement conditions.

\begin{figure*}[t]
\centering
\includegraphics[width=0.99\textwidth]{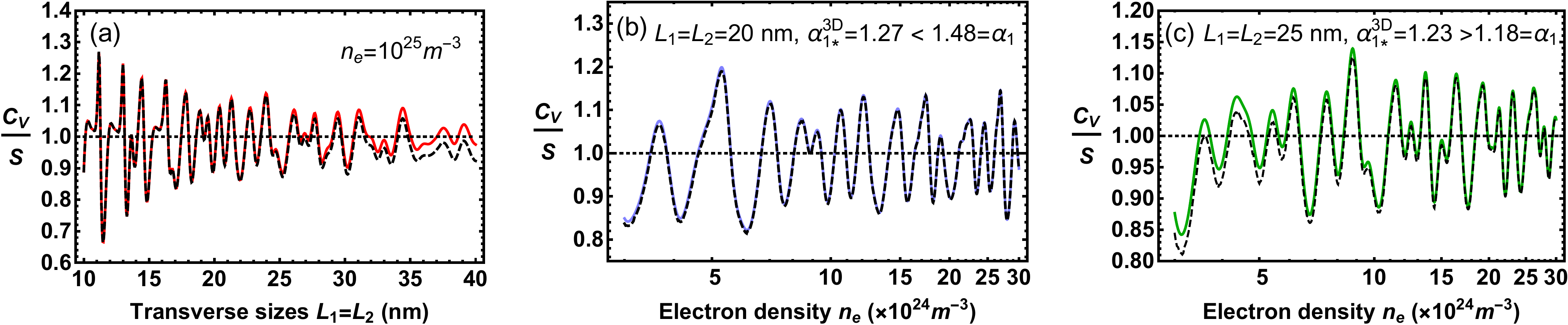}
\caption{(Color online) Oscillatory violation of entropy-heat capacity equivalence in degenerate limit depending on (a) size and (b), (c) density for a nanowire with 200 nm long at 5K temperature. Solid curves are the results obtained by using exact expressions, while dashed black curves are obtained by using HVM.}
\label{fig:pic8}
\end{figure*}

Very similar to the behavior of TOV, fully occupied or completely unoccupied states do not contribute to the entropy of the gas and contributions come only from the states around Fermi level, which are identified as HV states in HVM. Therefore, it is possible to use HVM also for the quantities which does not directly contain occupancy variance, as long as HV states are dominant over the quantity to be calculated like in entropy.

%%%        %%%%%%%%%%%%%%%%%%%%%%%%%%%%%%%%%%%%%%%%%%%%%%%%%%%%%%%%%%%        %%%
%%%        %%%%%%%%%%%%%%%%%%%%%%%% SECTION 5 %%%%%%%%%%%%%%%%%%%%%%%%        %%%
%%%        %%%%%%%%%%%%%%%%%%%%%%%%%%%%%%%%%%%%%%%%%%%%%%%%%%%%%%%%%%%        %%%

\section{Resemblance of total occupancy variance and density of states functions}
There is a direct relationship between DOS at Fermi level and TOV. It's well-known that, in strongly degenerate case, spin normalized distribution function nearly turns into a Heaviside step function $f(\tilde{\varepsilon})\approx g_s\Theta(\Lambda-\tilde{\varepsilon})$. Since the number of states below Fermi level $\Lambda$ is approximately equal to the number of Fermions per spin state, $N=\sum f(\tilde{\varepsilon})\approx g_s\sum\Theta(\Lambda-\tilde{\varepsilon})=NOS(\Lambda)$, derivative of $NOS(\Lambda)$ with respect to $\Lambda=\tilde{\varepsilon}_F$ gives $DOS(\Lambda)$ and then TOV is approximately equal to $DOS(\Lambda)$.
\begin{equation}
\Sigma_{D}^2=-\sum_{i_n}{\frac{\partial f}{\partial \tilde{\varepsilon}}}\approx g_s\sum_{i_n}{\delta(\Lambda-\tilde{\varepsilon})}=DOS(\Lambda).
\end{equation}
In Fig. 9, resemblance of TOV and DOS functions is shown for two different subbands with constant confinement parameters for various dimensions. Note that the results of lower dimensions are obtained by strongly confining a 3D domain in different directions. Even the lower dimensional domains are represented by triple sums where sums in strongly confined directions represent subbands. Hence, it's possible to directly obtain Eq. (6) by using continuous DOS functions at Fermi level $\tilde{\varepsilon}_F=\Lambda$ given in Table I. While HVM predicts oscillations quite good, DOS functions represents only the trend of the oscillations (Fig. 9) like Weyl representations of TOV functions in Fig. 2 and continuous TOV function in Fig. 3.

\begin{figure}[t]
\centering
\includegraphics[width=0.45\textwidth]{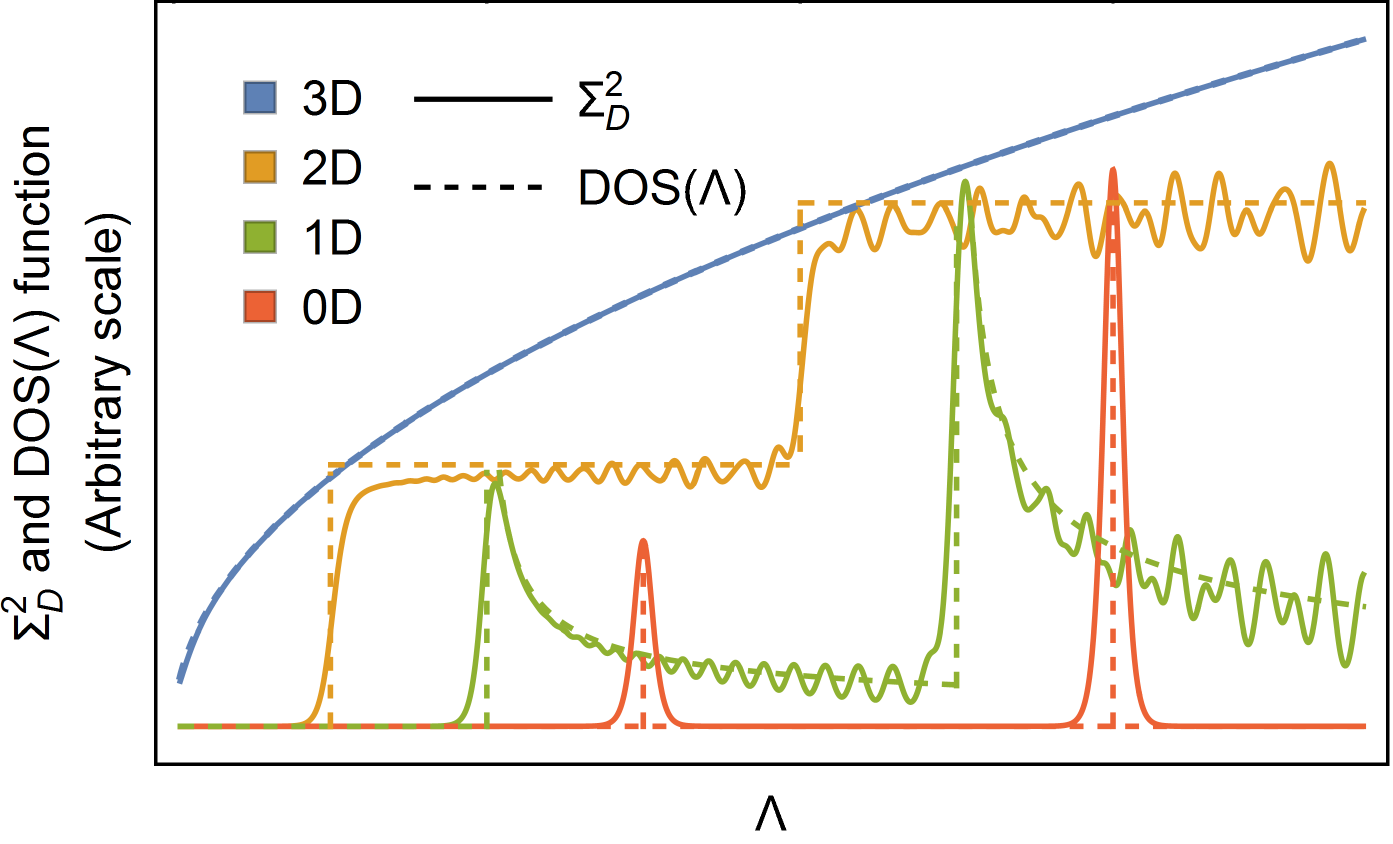}
\caption{(Color online) Similarity of TOV functions (solid curves) and continuous density of states functions at Fermi level (dashed curves) for various dimensions.}
\label{fig:pic9}
\end{figure}

Although the structure of two functions are very similar, due to thermal broadening at Fermi level there are small differences around the beginning of each subband especially for low dimensional cases. The origin of this resemblance is basically coming from the fact that occupancy variance function has a peakwise nature at Fermi level and vanishing behavior elsewhere, which is directly related with density of states at Fermi level.

%%%        %%%%%%%%%%%%%%%%%%%%%%%%%%%%%%%%%%%%%%%%%%%%%%%%%%%%%%%%%%%        %%%
%%%        %%%%%%%%%%%%%%%%%%%%%%%% SECTION 6 %%%%%%%%%%%%%%%%%%%%%%%%        %%%
%%%        %%%%%%%%%%%%%%%%%%%%%%%%%%%%%%%%%%%%%%%%%%%%%%%%%%%%%%%%%%%        %%%

\section{Conclusion}
In this article, investigation of size and density dependent quantum oscillations is done by proposing the HVM for the calculation of oscillatory quantities. Proposition of our model not only allows to calculate oscillatory thermodynamic or transport quantities in a more efficient way, but also offers a phase diagram which predict quantum oscillations clearly. Moreover, HVM introduces analytical conditions separating oscillatory and stationary regimes on the phase diagram.

Occupancy variance function appearing in many oscillatory quantities is examined in detail and the model is constructed on the half-vicinity of states around Fermi level. Two main factors determine the value of TOV: the number of states inside the half-vicinity shell and their proximities to the Fermi level. Confinement and degeneracy directly affect the distribution of the states around Fermi level besides the thickness of half-vicinity shell and so the number of states inside the shell.

The results show that HVM can be used as a reliable tool to predict and calculate quantum oscillations of a degenerate and confined ideal Fermi gas. The proposed phase diagram and analytical conditions for quantum oscillations evidently describe entirely stationary/oscillatory and crossover regions on degeneracy-confinement space. Consequently, our model and the phase diagram not only clarify the theoretical understanding of the quantum oscillations but also may help to determine the proper material parameters for experimental studies.

\bibliography{varef}
\bibliographystyle{unsrt}
\end{document}